\documentclass{article}

\usepackage{spconf,amsmath,graphicx}
\usepackage{url}
\usepackage{algorithmic}
\usepackage{algorithm}
\usepackage{amsfonts}

\newenvironment{itemize*}%
  {\begin{itemize}%
    \setlength{\itemsep}{0pt}%
    \setlength{\parskip}{0pt}}%
  {\end{itemize}}
  
\newcommand{\obs}{\text{obs}}
\newcommand{\D}{\mathcal{D}}

\newcommand{\BR}{\mathbb{R}}
\newcommand{\Cov}{\mathbb{C}\mathrm{ov}}
\newcommand{\N}{{\mathcal{N}}}

\newcommand{\GP}{\text{GP}}

\title{Randomized Maximum Likelihood via High-Dimensional Bayesian Optimization}
\twoauthors
 {Valentin Breaz\thanks{This work was partially funded by TotalEnergies.}}
	{Univ. Grenoble Alpes, Inria, CNRS, Grenoble INP*\\
	LJK, Grenoble, France}
 {Richard Wilkinson}
	{School of Mathematical Sciences\\
	University of Nottingham, UK}

\begin{document}
\maketitle
\begin{abstract}
Posterior sampling for high-dimensional Bayesian inverse problems is a common challenge in real-world applications.  Randomized Maximum Likelihood (RML) is an optimization based  methodology that gives samples from an approximation to the posterior distribution. We develop a high-dimensional Bayesian Optimization (BO) approach based on Gaussian Process (GP) surrogate models to solve the RML problem. We demonstrate the benefits of our approach in comparison to alternative optimization methods on a variety of synthetic and real-world Bayesian inverse problems, including medical and magnetohydrodynamics applications. 
\end{abstract}
\begin{keywords}
Bayesian Optimization, Gaussian Processes, Randomized Maximum Likelihood, Inverse Problems
\end{keywords}
\section{Introduction}
\label{sec:intro}

We consider Bayesian inverse problems, where the goal is to sample from the posterior distribution $p(x|\D)=\frac{p(\D|x)p(x)}{p(\D)}$
of the unknown parameters $x\in\BR^D$ given the observed data $\D\in\BR^m$.  Typically, the likelihood is assumed to be a Gaussian distribution $\D|x \sim\N_m(f(x),\Sigma_{\obs}),$ where $f(x):\BR^D\to\BR^m$ is known as the simulator; $f$ is often computationally expensive to evaluate and it usually models an underlying physical process. The covariance matrix $\Sigma_{\obs}$ describes the modelling and observational errors. Bayesian inverse problems are encountered in applications such as climate modelling \cite{Holden2018ABCFC}, medical imaging \cite{Dunlop2015TheBF}, and material sciences \cite{Iglesias2018BayesianII}.

Randomized Maximum Likelihood (RML) was introduced in \cite{kitanidis1995quasi, eage:/content/papers/10.3997/2214-4609.201406884}, as an approximate posterior sampling methodology. RML is formulated for the situation where the observations are subject to Gaussian distributed errors (i.e., a Gaussian likelihood) and where the prior distribution is also Gaussian: $x \sim\mathcal{N}_D(\mu, \Sigma)$. The algorithm proceeds by first perturbing the data and the prior mean, and then optimizing the unnormalised log-posterior using these perturbed values. See Algorithm \ref{CHalgorithm}.

\begin{algorithm}[H]
	\caption{Randomized Maximum Likelihood (RML)}
	\label{CHalgorithm}
	\begin{algorithmic}[0]
		\STATE $n_{RML}:$ number of samples required
		\FOR {$n\in [n_{RML}]$} 
			\STATE 1. Sample $\D_n\sim \N_m(\D, \Sigma_\obs)$ from the Gaussian likelihood
			\STATE 2. Sample $\mu_n\sim \N_D(\mu, \Sigma)$ from the Gaussian prior
			\STATE 3. Construct $O_n(x):=\log p(\D|x)p(x)$ w.r.t. the randomizations $(\D_n, \mu_n)$
			\begin{equation}\label{eq:eq1}
			 O_n(x)=\log\N_m(\D_n| f(x), \Sigma_\obs)+\log\N_D(x|\mu_n, \Sigma)
			\end{equation}
			\STATE 4. Obtain $x^\star_n$ as the maximizer $x^\star_n=\arg\max_x O_n(x)$.
		\ENDFOR
	\end{algorithmic}
\end{algorithm}

Here, we use the notation $[n_{RML}]=\{1, 2, \ldots, n_{RML}\}$. Thus, RML solves $n_{RML}$ optimization problems, each with a different objective function, $O_n(x)$. The resulting solutions $\{x^\star_n\}_{n=1}^{n_{RML}}$ are  approximate samples from the posterior distribution $p(x|\D)$. The samples are exact draws from the posterior if the simulator $f(x)$ is linear; see Appendix \ref{app:sec1} for a simple proof, or \cite{oliver1996conditional} for the original proof treating a more general case. Nonetheless, good practical performance has been observed for nonlinear (deep) neural network parametrized simulators  \cite{2020JCoPh.41309456T}; see also \cite{ba2022randomized} for a modification of RML for multi-modal posteriors with highly nonlinear simulators.

Instead of focusing on the accuracy of the approximate samples $\{x^\star_n\}_{n=1}^{n_{RML}}$ with respect to the true posterior, we address solving the optimization problems (\ref{eq:eq1}) efficiently in the challenging case of a high-dimensional input space $\mathbb{R}^D$. We focus on the specific scenario where the log-likelihood
\begin{align}\label{ll}
\log p(\D|x)\propto L(x):=-||\D-f(x)||_{\Sigma_\obs}^2
\end{align}
has a low-dimensional active subspace \cite{2015arXiv151000024C}, where we use the notational convention that $||y||^2_\Sigma=y^\top \Sigma^{-1} y$. In other words, we assume $L(x)\approx g(A^Tx)$, where $g:\BR^d\to\BR$ with $d\ll D$,  and $A\in\BR^{D\times d}$ is a semi-orthogonal matrix (i.e., $A^TA=I_d$); $A$ is known as the active subspace. As pointed out in \cite{2015arXiv151000024C}, many log-likelihoods used in inverse problems possess such a low-dimensional structure, e.g., in medical applications \cite{article_tyson}.

Bayesian Optimization (BO, i.e., finding $\arg\max_x O(x)$ for a generic objective function $O(x)$ using a Gaussian process surrogate model $g_\GP(x)\approx O(x)$) is based on a standard exploration-exploitation principle. Namely, an acquisition function based on $g_\GP(x)$ is used such that in the exploration phase, the target function $O(x)$ is explored globally, whereas in the exploitation phase, points $\tilde{x}$ that are likely to satisfy $\tilde{x}=\arg\max_x O(x)$ are sampled until the maximum is eventually found. For an introduction, see \cite{letham2020re} for BO, and \cite{10.5555/1162254} for Gaussian processes (GPs). In our setting, we deal with high-dimensional objectives $O(x)$ with an active subspace; however, we do not assume access to the active subspace. The most common solution is the use of random projections (embeddings) instead of the true low-dimensional active subspace; see Algorithm \ref{CHalgorithm3}, and \cite{wang2016bayesian} for a seminal work.

\begin{algorithm}[H]
	\caption{Standard high-dimensional BO algorithm with random embeddings for a generic objective function $O(\cdot)$}
	\label{CHalgorithm3}
	\begin{algorithmic}[0]
		\STATE $M:$ number of evaluations possible for a generic objective function $O(\cdot)$ given the computational budget;
		\STATE $d_e:$ chosen dimensionality of the embedding $R$;
		\STATE $R\in\mathbb{R}^{D\times d_e}:$ random embedding;
		\STATE $m_0:$ initial training points $\{y_i, O(Ry_i)\}_{i=1}^{m_0}$, $y_i\in\mathbb{R}^{d_e}$
		\FOR{$m\in\{m_0+1,\dots, M\}$}
			\STATE 1. Construct a GP approximation $O(Ry)\sim\GP$ using the available objective function evaluations  $\{y_i, O(Ry_i)\}_{i=1}^{m-1}$ 
			\STATE 2. Select $y_{m}=\arg\max_{y}a_m(y)$ as the maximizer of a BO acquisition function $a_m(y)$ for $O(Ry)\sim\GP$ 
			\STATE 3. Update the training data to $\{y_i, O(Ry_i)\}_{i=1}^{m}$.
		\ENDFOR
		\STATE Obtain $x_\star=Ry_{m_\star}$ as the maximizer 
		$$m_\star=\arg\max_{m} O(Ry_m),\,\,\,m\leq M.$$
	\end{algorithmic}
\end{algorithm}

Although there is an extensive literature for high dimensional BO (HD-BO) with random embeddings, which offers theoretical guarantees and good practical performance (see \cite{letham2020re} for a recent survey on the topic), there is no methodology designed for posterior sampling. Yet RML is a natural way to use HD-BO for high-dimensional posterior sampling. The closest related work to ours is \cite{articlehamdi}, where BO is compared with alternative gradient-free optimization methods for maximizing the log-likelihood in low-dimensional inverse problems. We extend their work by considering the multi-objective setting ($n_{RML}$ randomized log-likelihoods) in order to do posterior sampling, as well as considering high-dimensional inverse problems with high-dimensional priors. 

\textbf{Contributions.} We introduce a new methodology for high-dimensional posterior sampling via BO (Algorithm \ref{CHalgorithm4}); we propose a natural way to exploit the shared simulator $f(x)$ that is present in all of the objective functions (\ref{eq:eq1}), as well as an adjustment needed to incorporate a prior distribution without a low-dimensional structure. By using random embeddings, we sidestep the difficulty of estimating the active subspace, which might be impossible under tight computational budgets. We show that in the limited budget setting, our methodology usually outperforms alternative gradient-free optimization methods, according to a series of synthetic and real-world experiments. Finally, we show that the samples produced by our method are indeed close to `true' RML samples (collected via an infinite computational budget), whilst also covering well the high posterior density regions.

\section{Methodology}
\label{sec:methodology}

Firstly, we consider the setting of a uniform prior, $x\sim U[a_i, b_i]_{i=1}^D$, where $[a_i, b_i]_{i=1}^D:=[a_1, b_1]\times\cdots\times[a_D, b_D]$. In this case, the posterior distribution is proportional to the likelihood in the region of the prior support, and
using our active subspace assumption the RML objectives (\ref{eq:eq1}) become
\begin{equation}\label{eq:eq3}
O_n(x)\propto L_n(x):=\log\N_m(f(x)|\D_n, \Sigma_\obs)\approx g_n(A_n^Tx),
\end{equation}
where  the objective functions now need to be maximized over the prior support $[a_i, b_i]_{i=1}^D\subseteq\BR^D$. Note that we do not assume knowledge of the active subspaces $A_n\in\BR^{D\times d}$ or the low-dimensional link functions $g_n:\BR^d\to\BR$.

We propose our procedure HD-BO-RML in Algorithm \ref{CHalgorithm4}. Following the suggestion of \cite{wang2016bayesian}, we use a collection of $K$ interleaved random embeddings.  Algorithm \ref{CHalgorithm4} is based on a cyclic pass through all the objective functions (\ref{eq:eq3}), where for every random embedding $k\in[K]$, the simulations $\{(y^k_m, f(R_ky^k_m))\}_{m=1}^n$ collected up to some iteration $n$ will serve as the basis for the training set used in the next iteration, where the objective function $O_{(n+1)'}(x)$ is considered, with $(n+1)':=n+1 \mod n_{RML}$. 

Given a budget of $N$ simulator evaluations, our procedure performs $T:=N/{n_{RML}}$ HD-BO iterations for each objective. If $N$ (and thus $T$) is large enough, then HD-BO-RML benefits from the same theoretical guarantees as a (vanilla) HD-BO algorithm for each objective function (e.g., Theorem 11 in \cite{wang2016bayesian}). However, instead of performing the $T$ HD-BO iterations independently for each objective, we further exploit that the simulator function $f(x)$ is present in all the objectives $L_n(x):=\log\mathcal{N}(f(x)|\mathcal{D}_n, \Sigma_{obs})$. Namely, with one HD-BO iteration for $L_j(x)$, where some point $y^k_j$ is selected by maximizing the acquisition function, we can use $(y^k_j, f(R_ky^k_j))$ to obtain (training) data for all the objectives. The next HD-BO iteration for $L_j(x)$ will come after $n_{RML}-1$ steps, and instead of having only one training point $(y^k_j, L_j(R_ky^k_j))$, we will have $n_{RML}-1$ extra points $\{(y^k_n, L_j(R_ky^k_n))\}_{n=j+1}^{n_{RML}+j-1}$ obtained from all the other objectives during the cyclic pass. 

What we tend to observe in experiments is that during the first few HD-BO iterations (exploration phase), the selected points $(y^k_n, f(R_ky^k_n))$ are informative of the overall landscape of the posterior space (i.e., they separate the low-posterior density regions from the high-posterior density regions), whereas training points selected during the exploitation phase are informative for the high-posterior density regions, and thus are beneficial for every RML objective.

\begin{algorithm}[H]
 	\caption{HD-BO-RML (Uniform or Gaussian priors; the additional steps for Gaussian priors are shown in parentheses)}
	\label{CHalgorithm4}
	\begin{algorithmic}[0]
		\STATE $N:$ max possible number of evaluations of $f(\cdot)$; 
		\STATE $d_e:$ choice of embedding dimensionality;
		\STATE $R_1, \ldots, R_K\in\mathbb{R}^{D\times d_e}:$ random embeddings;
		\STATE $n_0\times K$ initial points: $\{y^k_i, f(R_ky^k_i)\}_{i=1}^{n_0}$, with $y^k_i\in\mathbb{R}^{d_e}$, $k\in[K]$;
        \FOR {$k\in\{1,\dots, K\}$}
		    \FOR{$n\in\{n_0+1,\dots, \lfloor N/K\rfloor\}$
		    ({$n\in\{n_0+1,\dots, \lfloor N/2K\rfloor\}$ in case of a Gaussian prior})}
    			\STATE 1. Let $n':= n \mod n_{RML}$
    			\STATE 2. Construct a GP approximation to $L_{n'}(R_ky)$ using the existing simulations  $\{y^k_i, f(R_ky^k_i)\}_{i=1}^{n-1}$ 
    			\STATE 3. Select $y^k_{n}=\arg\max_{y}a^k_n(y)$ as the maximizer of a BO acquisition function using the GP approximation
    			\STATE 4. Perform $f(R_ky^k_n)$ and update the shared simulation ensemble (i.e., training set) to $\{y^k_i, f(R_ky^k_i)\}_{i=1}^{n}$
    			\STATE (Gaussian prior) {5G. Perform local optimization in $\mathbb{R}^D$ around $x_0=R_ky^k_n$ with respect to the prior $p_{n'}(x)=\mathcal{N}_D(x|\mu_{n'}, \Sigma)$ 
			    }
		    \ENDFOR
		\ENDFOR
		\FOR {$n\in\{1,\dots, n_{RML}\}$}
    		\STATE (Uniform prior) 1U. Obtain $x^\star_n=R_{k_\star}y^{k_\star}_{m_\star}$ as the maximizer 
    		$$k_\star, m_\star=\arg\max_{k, m} O_n(R_ky^k_m),\,k\in[K],m\leq\lfloor N/K\rfloor$$
    		\STATE { (Gaussian prior) 1G. Writing $z_m^k$ for the local maxima from step 5G. above, obtain $x^\star_n=z^{k_\star}_{m_\star}$ via
    		$$k_\star, m_\star=\arg\max_{k, m} O_n(z_m^k),\,k\in[K],m\leq\lfloor N/2K\rfloor$$}
		\ENDFOR
	\end{algorithmic}
\end{algorithm}

Using a Gaussian prior distribution, $x\sim\N_D(\mu, \Sigma)$, rather than a uniform prior, requires a modification of the algorithm. Using again the active subspace assumption, the objective functions (\ref{eq:eq1}) become:
\[O_n(x)=L_n(x)+\log p_n(x)\approx g_n(A_n^Tx)+\log\N_D(x|\mu_n, \Sigma).\] Due to the potential lack of a low-dimensional (linear) subspace in the prior, as for example if $x\sim\N(0, I_D)$, running HD-BO by directly modelling $O_n(Ry)$ as a GP might be unsatisfactory, and hence we cannot simply use the algorithm from the uniform prior case. Steps 5G and 1G in Algorithm \ref{CHalgorithm4} propose a heuristic solution.
While we keep performing HD-BO with respect to the log-likelihood $L_{n'}(x)\approx g_{n'}(A_{n'}^Tx)$ as in the uniform prior case,
we try to increase the prior value for the points $y^k_{n}=\arg\max_{y}a^k_n(y)$ of potentially high-likelihood $L_{n'}(R_ky^k_n)$ by carefully performing local optimization with respect to the prior $p_{n'}(x)=\mathcal{N}_D(x|\mu_{n'}, \Sigma)$ in the high-dimensional space $\mathbb{R}^D$, starting from $x_0:=R_ky^k_n$; the resulting local maxima $z_n^k$ try to achieve $L_{n'}(z_n^k)+p_{n'}(z^k_n)>L_{n'}(R_ky^k_n)+p_{n'}(R_ky^k_n)$ (i.e., $O_{n'}(z_n^k)>O_{n'}(R_ky^k_n)$). This inequality was indeed observed in our experiments; during the exploitation stage, $z_n^k$ managed to achieve a trade-off between high-likelihood and high-prior. Note that the local maximization (step 5G) does not require simulator evaluations, and any local optimization method can be applied without any significant costs. 

\section{Experiments}
\label{sec:exp}

We now give empirical results showing that the proposed algorithm performs well in comparison with competing methods in a variety of synthetic and real-world Bayesian inverse problems. We use four simulators from the Active Subspaces github page maintained by \cite{asdatasets}:
\begin{itemize*}
	\item \textbf{Elliptic-PDE} $f:\BR^{100}\to\BR^{7}$. 
	This is a widely used exemplar for Bayesian inverse problems.
	\item \textbf{Ebola}: $f:\BR^{8}\to\BR$, an 8-parameter dynamical system model for the spread of Ebola in Liberia \cite{diaz2018modified}.
	\item \textbf{MHD}: $f:\BR^{5}\to\BR$, a 5-parameter magnetohydrodynamics power generation model \cite{glaws2017dimension}. 
	\item \textbf{HIV long-term model}: $f:\BR^{27}\to\BR$ is the cell count at time $t=24$ days \cite{article_tyson}.
\end{itemize*}

In each case, the  log-likelihood has a low-dimensional active subspace $L(x)\approx g(A^Tx)$ with $g:\BR\to\BR$ for the HIV model, and $g:\BR^2\to\BR$ for the other simulators. We do not assume knowledge of $A$. We use a standard Gaussian prior $x\sim\N_{100}(0, I)$ for the elliptic-PDE, and a uniform prior distribution for the other simulators. For each experiment, we use RML to approximately sample from the posterior distribution, comparing our high dimensional Bayesian optimization (HD-BO) approach \textbf{HD-BO-RML} against  the following gradient-free optimization methods: \textbf{BOBYQA} \cite{article_powell}, a trust-region algorithm, used for method comparison versus HD-BO in \cite{NEURIPS2019_6c990b7a}; \textbf{CMA-ES} \cite{hansen2019pycma},  an evolution strategy algorithm, used for method comparison in  \cite{NEURIPS2019_6c990b7a, letham2020re}; \textbf{NSGA-II} \cite{996017}, an evolutionary algorithm for multi-objective optimization, used for method comparison versus HD-BO in \cite{daulton2022multi}; and \textbf{random design} \cite{wang2016bayesian}, a standard baseline for comparison versus BO. We measure performance of the different methods by comparing the mean return,  defined as: $1/{n_{RML}}\sum_{n=1}^{n_{RML}}O_n(x^\star_n)$, where $x^\star_n$ is the approximate maximizer of $O_n(x)$ as selected by the different optimization methods considered.

For HD-BO-RML, we use the Gaussian Process Upper Confidence Bound (GP-UCB) acquisition function 
\cite{10.5555/3104322.3104451}, i.e., $a^k_n(y)=\mu^k_n(y)+\beta\sigma^k_n(y),$
where $\mu^k_n(y)$ is the GP predictive mean which tries to approximate the log-likelihood $L_{n'}(R_ky$) ($n':=n \mod n_{RML}$), and $\sigma^k_n(y)$ represents the GP predictive uncertainty. This acquisition function has been studied in the HD-BO literature with random embeddings \cite{wang2016bayesian}. We use a standard squared-exponential covariance function (RBF kernel), i.e.,$\Cov\large(O(Ry_1), O(Ry_2)\large):=o^2\exp(-\frac{||y_1-y_2||^2}{2l^2})$ for each GP model, where $o$ and $l$  are trainable hyperparameters known as the outputscale and lengthscale, respectively. For the random embeddings, $R_k$, each row is sampled independently from the uniform distribution on the unit hypersphere $\mathbb{S}^{d_e-1}$, as suggested in \cite{letham2020re}. We use $K=10$ random embeddings of low dimensionality $d_e=d+1$  and $n_0=5$ initial points (where $d$ is the dimension of the active subspace).

\subsection{Results}
\label{sec:results}

\begin{figure}[htb]
\begin{minipage}[b]{.48\linewidth}
  \centering
  \centerline{\includegraphics[width=4.0cm]{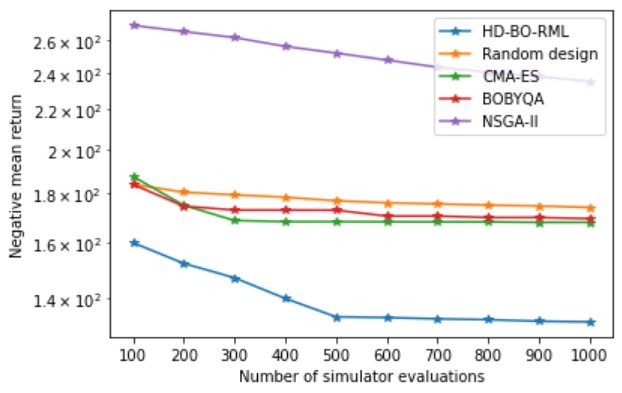}}
  \centerline{Elliptic PDE}\medskip
\end{minipage}
\hfill
\begin{minipage}[b]{0.48\linewidth}
  \centering
  \centerline{\includegraphics[width=4.0cm]{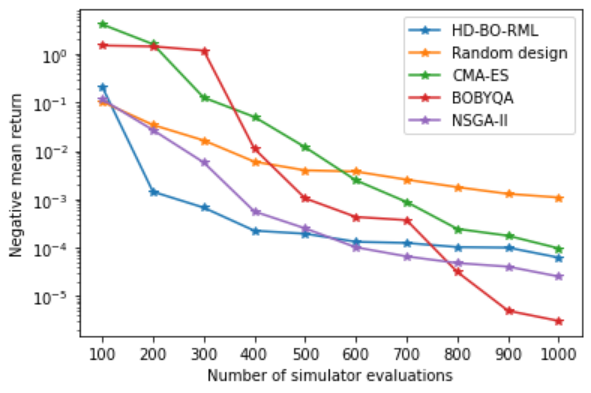}}
  \centerline{Ebola}\medskip
\end{minipage}
\begin{minipage}[b]{.48\linewidth}
  \centering
  \centerline{\includegraphics[width=4.0cm]{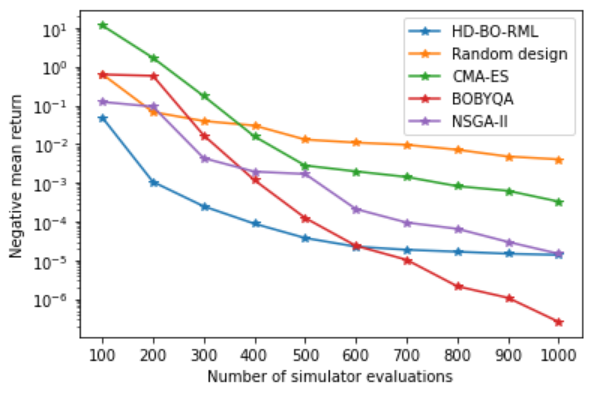}}
  \centerline{MHD}\medskip
\end{minipage}
\hfill
\begin{minipage}[b]{0.48\linewidth}
  \centering
  \centerline{\includegraphics[width=4.0cm]{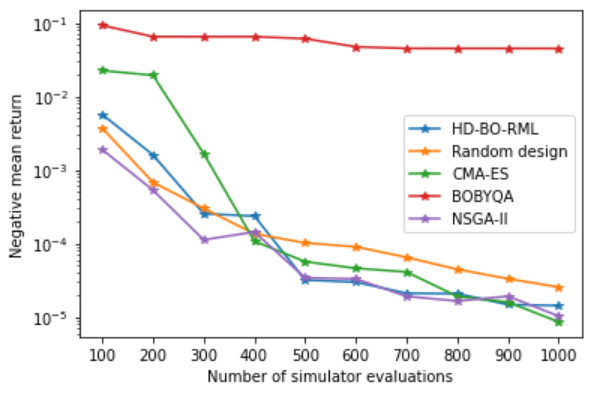}}
  \centerline{HIV}\medskip
\end{minipage}
\caption{Negative mean returns: $-1/{n_{RML}}\sum_{n=1}^{n_{RML}}O_n(x^\star_n)$ (lower is better) versus computational budget for each method}
\label{fig:method_comparison}
\end{figure}

Figure \ref{fig:method_comparison} shows the performance of each algorithm for different computational budgets, averaged over 5 trials.  Each trial has a budget of $N=1000$ simulations in order to find $n_{RML}=20$ samples. We can see that HD-BO-RML has the best performance of all the methods for small budgets (i.e., small $N$), and has comparable performance to other methods with larger budgets; in each experiment, HD-BO-RML is the best performing method across the full range of computational budgets for at least one of the 5 trials (and in all the 5 trials for the high-dimensional Elliptic PDE with $D=100$). With a fixed budget, the performance for HD-BO-RML should remain relatively stable as the number $n_{RML}$ of RML samples required increases (since exploration/exploitation will still be performed jointly); CMA-ES and BOBYQA will perform significantly worse due to fewer iterations and no data sharing between objectives, while NSGA-II is known to struggle for a large number of objectives \cite{10.5555/1762545.1762607}. 

\begin{figure}[htb]
\begin{minipage}[b]{.3\linewidth}
  \centering
  \centerline{\includegraphics[width=2.7cm]{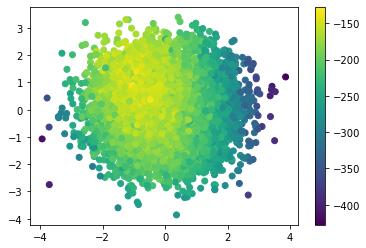}}
  \label{fig:figpdeeee}
  \centerline{PDE}\medskip
\end{minipage}
\begin{minipage}[b]{.3\linewidth}
  \centering
  \centerline{\includegraphics[width=2.7cm]{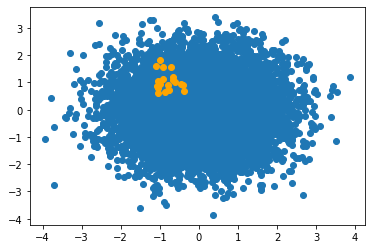}}
  \centerline{RML oracle}\medskip
\end{minipage}
\begin{minipage}[b]{.3\linewidth}
  \centering
  \centerline{\includegraphics[width=2.7cm]{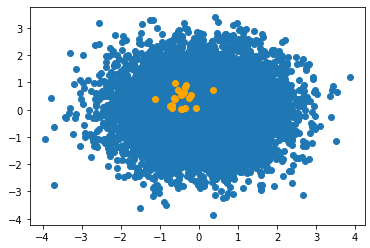}}
  \centerline{HDBO-RML}\medskip
\end{minipage}
\begin{minipage}[b]{.3\linewidth}
  \centering
  \centerline{\includegraphics[width=2.7cm]{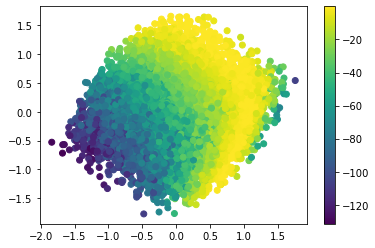}}
  \centerline{Ebola}\medskip
\end{minipage}
\begin{minipage}[b]{.3\linewidth}
  \centering
  \centerline{\includegraphics[width=2.7cm]{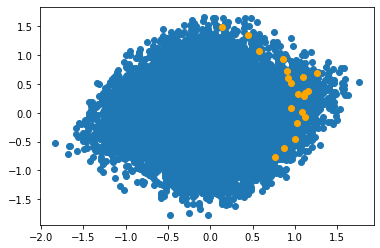}}
  \centerline{RML oracle}\medskip
\end{minipage}
\begin{minipage}[b]{.3\linewidth}
  \centering
  \centerline{\includegraphics[width=2.7cm]{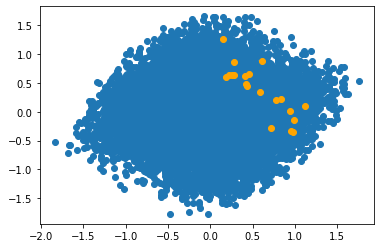}}
  \centerline{HDBO-RML}\medskip
\end{minipage}
\begin{minipage}[b]{.3\linewidth}
  \centering
  \centerline{\includegraphics[width=2.7cm]{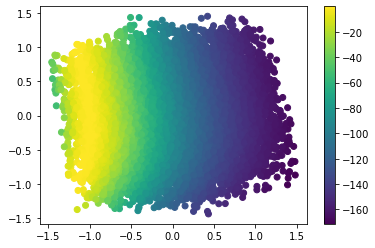}}
  \centerline{MHD}\medskip
\end{minipage}
\hfill
\begin{minipage}[b]{.3\linewidth}
  \centering
  \centerline{\includegraphics[width=2.7cm]{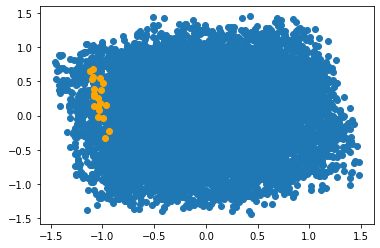}}
  \centerline{RML oracle}\medskip
\end{minipage}
\hfill
\begin{minipage}[b]{.3\linewidth}
  \centering
  \centerline{\includegraphics[width=2.7cm]{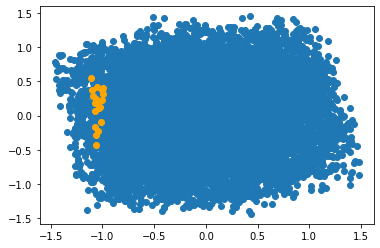}}
  \centerline{HDBO-RML}\medskip
\end{minipage}
\begin{minipage}[b]{.3\linewidth}
  \centering
  \centerline{\includegraphics[width=2.7cm]{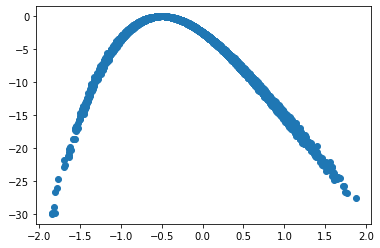}}
  \centerline{HIV}\medskip
\end{minipage}
\hfill
\begin{minipage}[b]{.3\linewidth}
  \centering
  \centerline{\includegraphics[width=2.7cm]{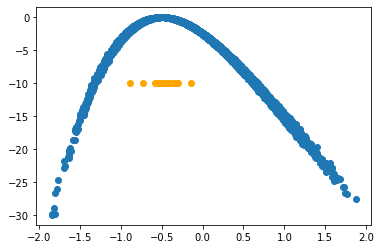}}
  \centerline{RML oracle}\medskip
\end{minipage}
\hfill
\begin{minipage}[b]{.3\linewidth}
  \centering
  \centerline{\includegraphics[width=2.7cm]{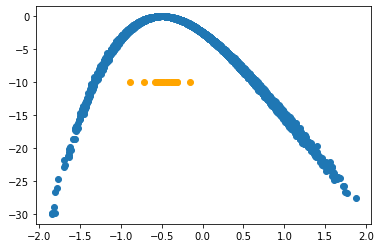}}
  \centerline{HDBO-RML}\medskip
\end{minipage}
\caption{Posterior landscape in the active subspace (left), oracle RML samples (middle), HD-BO-RML samples (right). RML samples are shown in orange, with prior samples in blue.}
\label{fig:posterior_visualization}
\end{figure}

As  in \cite{2015arXiv151000024C}, we can visualize each posterior landscape $p(x|\D)$ in the active subspace. The plots in the left column of Figure \ref{fig:posterior_visualization} show 10000 prior samples $x_i\sim p(x)$  projected into the true two-dimensional (one-dimensional for HIV) active subspace $A$, coloured by their unnormalized log-posterior density, i.e., $\large\{A^Tx_i, \log p(\D|x_i)+\log p(x_i)\large\large\}_{i=1}^{10000}.$ We use an oracle with access to an unlimited amount of simulator evaluations to obtain the `true' $n_{RML}=20$ RML samples; the resulting samples are represented in orange in the middle column of Figure \ref{fig:posterior_visualization}. We see that the RML samples cover the high posterior density regions well in all the experiments. Finally, the right column of Figure \ref{fig:posterior_visualization} shows the analogous samples obtained from using HD-BO-RML (instead of the oracle optimizer), averaged over the 5 trials. We see that the HD-BO-RML samples are close to the oracle samples in the active subspace representation.

\section{Discussion}
\label{sec:discussion}

Our focus has been on the challenging and commonly occurring problem  of posterior sampling using a complex simulator, but with limited computational budget and no access to gradient information. We have introduced an RML approach based on high-dimensional Bayesian optimization that outperforms  competing gradient-free optimization methods when there are tight computational budget constraints. Here we assumed the active subspace is unknown, but if it is known, the same BO approach can be used with the true active subspace instead of the random embeddings.

\bibliographystyle{IEEEbib}
\bibliography{main}

\begin{thebibliography}{10}

\bibitem{Holden2018ABCFC}
Philip~B. Holden, Neil~Robert Edwards, J.~S. Hensman, and Richard~D. Wilkinson,
\newblock ``Abc for climate: Dealing with expensive simulators,''
\newblock {\em Handbook of Approximate Bayesian Computation}, 2018.

\bibitem{Dunlop2015TheBF}
Matthew~M. Dunlop and Andrew~M. Stuart,
\newblock ``The bayesian formulation of eit: Analysis and algorithms,''
\newblock {\em arXiv: Probability}, 2015.

\bibitem{Iglesias2018BayesianII}
Marco~A. Iglesias, Minho Park, and Michael~V. Tretyakov,
\newblock ``Bayesian inversion in resin transfer molding,''
\newblock {\em Inverse Problems}, 2018.

\bibitem{kitanidis1995quasi}
Peter~K Kitanidis,
\newblock ``Quasi-linear geostatistical theory for inversing,''
\newblock {\em Water resources research}, vol. 31, no. 10, pp. 2411--2419, 1995.

\bibitem{eage:/content/papers/10.3997/2214-4609.201406884}
D.~S.~Oliver, N.~He, and A.~C.~Reynolds,
\newblock ``Conditioning permeability fields to pressure data,''
\newblock {\em European Association of Geoscientists \& Engineers}, 1996.

\bibitem{oliver1996conditional}
Dean~S Oliver,
\newblock ``On conditional simulation to inaccurate data,''
\newblock {\em Mathematical Geology}, vol. 28, no. 6, pp. 811--817, 1996.

\bibitem{2020JCoPh.41309456T}
Meng {Tang}, Yimin {Liu}, and Louis~J. {Durlofsky},
\newblock ``{A deep-learning-based surrogate model for data assimilation in dynamic subsurface flow problems},''
\newblock {\em Journal of Computational Physics}, vol. 413, July 2020.

\bibitem{ba2022randomized}
Yuming Ba, Jana de~Wiljes, Dean~S Oliver, and Sebastian Reich,
\newblock ``Randomized maximum likelihood based posterior sampling,''
\newblock {\em Computational Geosciences}, pp. 1--23, 2022.

\bibitem{2015arXiv151000024C}
Paul~G. {Constantine}, Carson {Kent}, and Tan {Bui-Thanh},
\newblock ``{Accelerating MCMC with active subspaces},''
\newblock {\em arXiv e-prints}, p. arXiv:1510.00024, Sept. 2015.

\bibitem{article_tyson}
Tyson Loudon and Stephen Pankavich,
\newblock ``Mathematical analysis and dynamic active subspaces for a long term model of hiv,''
\newblock {\em Mathematical Biosciences and Engineering}, vol. 14, 04 2016.

\bibitem{letham2020re}
Ben Letham, Roberto Calandra, Akshara Rai, and Eytan Bakshy,
\newblock ``Re-examining linear embeddings for high-dimensional bayesian optimization,''
\newblock {\em Advances in neural information processing systems}, vol. 33, 2020.

\bibitem{10.5555/1162254}
Carl~Edward Rasmussen and Christopher K.~I. Williams,
\newblock {\em Gaussian Processes for Machine Learning (Adaptive Computation and Machine Learning)},
\newblock The MIT Press, 2005.

\bibitem{wang2016bayesian}
Ziyu Wang, Frank Hutter, Masrour Zoghi, David Matheson, and Nando De~Feitas,
\newblock ``Bayesian optimization in a billion dimensions via random embeddings,''
\newblock {\em Journal of Artificial Intelligence Research}, vol. 55, 2016.

\bibitem{articlehamdi}
Hamidreza Hamdi, Ivo Couckuyt, Mario Costa~Sousa, and Tom Dhaene,
\newblock ``Gaussian processes for history-matching: application to an unconventional gas reservoir,''
\newblock {\em Computational Geosciences}, vol. 21, 04 2017.

\bibitem{asdatasets}
P.~Constantine and R.~Howard,
\newblock ``Active subspaces data sets,'' \url{https://github.com/paulcon/as-data-sets}.

\bibitem{diaz2018modified}
Paul Diaz, Paul Constantine, Kelsey Kalmbach, Eric Jones, and Stephen Pankavich,
\newblock ``A modified seir model for the spread of ebola in western africa and metrics for resource allocation,''
\newblock {\em Applied mathematics and computation}, vol. 324, pp. 141--155, 2018.

\bibitem{glaws2017dimension}
Andrew Glaws, Paul~G Constantine, John~N Shadid, and Timothy~M Wildey,
\newblock ``Dimension reduction in magnetohydrodynamics power generation models: Dimensional analysis and active subspaces,''
\newblock {\em Statistical Analysis and Data Mining: The ASA Data Science Journal}, vol. 10, no. 5, pp. 312--325, 2017.

\bibitem{article_powell}
M.~Powell,
\newblock ``A view of algorithms for optimization without derivatives,''
\newblock {\em Mathematics TODAY}, vol. 43, 01 2007.

\bibitem{NEURIPS2019_6c990b7a}
David Eriksson, Michael Pearce, Jacob Gardner, Ryan~D Turner, and Matthias Poloczek,
\newblock ``Scalable global optimization via local bayesian optimization,''
\newblock in {\em Advances in Neural Information Processing Systems}, H.~Wallach, H.~Larochelle, A.~Beygelzimer, F.~d\textquotesingle Alch\'{e}-Buc, E.~Fox, and R.~Garnett, Eds., 2019, vol.~32.

\bibitem{hansen2019pycma}
Nikolaus Hansen, Youhei Akimoto, and Petr Baudis,
\newblock ``{CMA-ES/pycma} on {G}ithub,'' Zenodo, DOI:10.5281/zenodo.2559634, Feb. 2019.

\bibitem{996017}
K.~Deb, A.~Pratap, S.~Agarwal, and T.~Meyarivan,
\newblock ``A fast and elitist multiobjective genetic algorithm: Nsga-ii,''
\newblock {\em IEEE Transactions on Evolutionary Computation}, vol. 6, no. 2, pp. 182--197, 2002.

\bibitem{daulton2022multi}
Samuel Daulton, David Eriksson, Maximilian Balandat, and Eytan Bakshy,
\newblock ``Multi-objective bayesian optimization over high-dimensional search spaces,''
\newblock in {\em Uncertainty in Artificial Intelligence}. PMLR, 2022.

\bibitem{10.5555/3104322.3104451}
Niranjan Srinivas, Andreas Krause, Sham Kakade, and Matthias Seeger,
\newblock ``Gaussian process optimization in the bandit setting: No regret and experimental design,''
\newblock in {\em Proceedings of the 27th International Conference on International Conference on Machine Learning}, Madison, WI, USA, 2010, ICML'10, p. 1015–1022, Omnipress.

\bibitem{10.5555/1762545.1762607}
Mario K\"{o}ppen and Kaori Yoshida,
\newblock ``Substitute distance assignments in nsga-ii for handling many-objective optimization problems,''
\newblock in {\em Proceedings of the 4th International Conference on Evolutionary Multi-Criterion Optimization}, Berlin, Heidelberg, 2007, EMO'07, p. 727–741, Springer-Verlag.

\end{thebibliography}

% \clearpage
\appendix

\section{Proof for the linear simulator case}\label{app:sec1}

In case of a linear simulator $f(x):=Hx$, Gaussian prior distribution $x\sim\N(\mu, \Sigma)$, and Gaussian likelihood $\D|x \sim\N(Hx,\Sigma_{\obs})$, the posterior distribution $p(x|\D)$ is also Gaussian: $$x | \D \sim \N(m, V),$$
where
\begin{align*}
V&=(\Sigma^{-1} + H^\top \Sigma_{\obs}^{-1} H)^{-1} \\
\mbox{ and } \quad m &= V(\Sigma^{-1}\mu + H^\top \Sigma_{\obs}^{-1}\D).
\end{align*}

To see why Randomized Maximum Likelihood (RML) from Algorithm \ref{CHalgorithm} produces exact samples from the posterior distribution in this case, we first recall the RML objective functions from (\ref{eq:eq1}):

$$O_n(x):=\log\N_m(Hx|\D_n, \Sigma_\obs)+\log\N_D(x|\mu_n, \Sigma),$$

where $\mu_n:=\mu+\epsilon_n$ and $\D_n:=\D+\eta_n$, with $\epsilon_n\sim\N(0, \Sigma)$ and $\eta_n\sim\N(0, \Sigma_{\obs})$ independently.

Differentiating with respect to $x$ gives
$$\nabla O_n(x) = 2\Sigma^{-1}x - 2\Sigma^{-1}\mu_n +2H^\top \Sigma_{\obs}^{-1} Hx - 2 H^\top \Sigma_{\obs}^{-1}\D_n$$
and setting this equal to zero and rearranging gives
\begin{align*}
x &= (\Sigma^{-1}+H^\top \Sigma_{\obs}^{-1} H)^{-1}(\Sigma^{-1}\mu_n + H^\top \Sigma_{\obs}^{-1}\D_n)\\
&= m+V(\Sigma^{-1}\epsilon_n + H^\top \Sigma_\obs^{-1}\eta_n).
\end{align*}

The distribution of $x$ can then easily be seen to be 
\begin{align*}
x &\sim \N(m, V(\Sigma^{-1}\Sigma\Sigma^{-1} + H^\top \Sigma_{\obs}^{-1}\Sigma_{\obs}\Sigma_{\obs}^{-1}H)V^\top)\\
&= \N(m,V),
\end{align*}
i.e., $x$ is a sample from the true posterior distribution.

\end{document}